\documentclass[12pt]{article}
\pdfoutput=1
\usepackage{latexsym}\usepackage{epsfig,amssymb,euscript}
\usepackage{amsmath} \usepackage{bbm,slashed}
\usepackage{color}
\usepackage{ulem}
\usepackage{cite}

\newcommand{\be}{\begin{equation}}
\newcommand{\ee}{\end{equation}}
\newcommand{\bea}{\begin{eqnarray}}
\newcommand{\eea}{\end{eqnarray}}

\newcommand{\nn} {\nonumber}

\catcode`@=12

\numberwithin{equation}{section}

\begin{document}
\begin{titlepage}

\begin{flushright}
SISSA 21/2012/EP
\end{flushright}
\bigskip
\def\thefootnote{\fnsymbol{footnote}}

\begin{center}
\vskip -10pt
{\LARGE
{\bf
Exploring Holographic \\
\vspace{0.30in}
General Gauge Mediation
}
}
\end{center}

\bigskip
\begin{center}
{\large
Riccardo Argurio,$^{1,2}$ 
Matteo Bertolini,$^{3,4}$ 
Lorenzo Di Pietro,$^3$  \\
\vskip 5pt
Flavio Porri$^3$ and
Diego Redigolo$^{1,2}$}

\end{center}

\renewcommand{\thefootnote}{\arabic{footnote}}

\begin{center}
$^1${Physique Th\'eorique et Math\'ematique \\
Universit\'e Libre de Bruxelles, C.P. 231, 1050 Bruxelles, Belgium\\}
$^2${International Solvay Institutes, Brussels, Belgium}\\
$^3$ {SISSA and INFN - Sezione di Trieste\\
Via Bonomea 265; I 34136 Trieste, Italy\\}
$^4$ {International Centre for Theoretical Physics (ICTP)\\
Strada Costiera 11; I 34014 Trieste, Italy}


\end{center}

\noindent
\begin{center} {\bf Abstract} \end{center}

We  study models of gauge mediation with
strongly coupled hidden sectors, employing  
a hard wall background as an holographic dual description. The structure
of the soft spectrum depends crucially on the boundary conditions one
imposes on bulk fields at the IR wall. Generically, vector
and fermion correlators have poles at zero momentum, leading to gauge mediation by massive
vector messengers and/or generating Dirac gaugino masses. Instead, non-generic choices of
boundary conditions let one cover all of GGM parameter space.
Enriching the background with R-symmetry breaking scalars, the SSM soft term
structure becomes more constrained and similar to previously studied
top-down models, while retaining the more analytic control the 
present bottom-up approach offers.
 
\vspace{1.6 cm}
\vfill

\end{titlepage}

\setcounter{footnote}{0}

\section{Introduction}

We have started in \cite{Argurio:2012cd} to investigate with
holographic methods \cite{Maldacena:1997re,Witten:1998qj,Gubser:1998bc,Bianchi:2001de,Bianchi:2001kw,Skenderis:2002wp} strongly
coupled hidden sectors with the aim of extracting data suitable for
models of gauge mediation of supersymmetry breaking, using the formalism of General Gauge Mediation (GGM) \cite{Meade:2008wd}. We have
laid out the general procedure one should follow, and applied it to a class of non-supersymmetric Asymptotically AdS (AAdS) backgrounds 
which arise as solutions of ten-dimensional type IIB supergravity  \cite{Kehagias:1999tr,Gubser:1999pk}. As such,
these backgrounds represent a top-down approach where the supersymmetry
breaking theory is found directly in a possibly string-derived and 
UV-complete set up  (see \cite{Benini:2009ff,McGuirk:2009am,McGuirk:2011yg,Skenderis:2012bs} for
other works in the same direction). 

The two rather general results we have obtained there can be summarized as follows:
\begin{itemize}
\item Pure dilatonic backgrounds describe, holographically, supersymmetry breaking hidden sectors where gauginos cannot acquire a Majorana mass. Rather, 
they can get Dirac masses, due to direct coupling to R-charged massless fermions, which arise quite generically in the low energy spectrum of the strongly coupled hidden 
sector. As such, the supersymmetric Standard Model (SSM) soft spectrum 
turns out to be similar to that of gaugino mediation models.
\item Backgrounds where scalars whose field theory dual operators are charged under the R-symmetry get a non-trivial profile, do instead generate gaugino Majorana masses. In such situations, no R-charged massless fermions show-up in the hidden sector, and hence there is no Dirac-like contribution to gaugino masses. Here the SSM soft spectrum looks democratic between gauginos and sfermions, and hence 
similar to that of minimal gauge mediation models.     
\end{itemize}

A natural question one might ask is whether these results are generic
for models admitting holographic duals, or whether they strongly
depend on the specific background we have been considering. More
generally, how large a portion of the GGM parameter space can
holographic models of gauge mediation cover? Are there any restrictions? These questions are difficult to answer within fully-fledged top-dow models, given also the poor number of concrete and sufficiently explicit string theory supersymmetry breaking solutions available in the literature.  

In the present paper we will focus on a complementary, bottom-up approach. We will consider simple backgrounds which do not have
necessarily a completion in string theory, but which have the advantage to let one 
have more flexibility and more analytical power. This will enable us 
to answer the above questions, postponing to future studies the possibility that the extra requirement of a full string embedding might eventually restrict further the region of GGM parameter space  one can cover via holographic models. 

We will keep considering AAdS backgrounds, and, without loss of generality, focus on a hidden sector global $U(1)$ symmetry. This is the symmetry one should    
weakly gauge in order to couple the (supersymmetry breaking) hidden sector dynamics to the SSM.  
We will focus on a simple hard wall model \cite{Polchinski:2001tt, BoschiFilho:2002vd, Erlich:2005qh} which
consists of a pure AdS background with an IR cut-off in the bulk. The
boundary conditions one imposes on the fluctuating fields at the IR
cut-off are the only freedom one has in this set up. We will 
parameterize the structure of 
the SSM soft spectrum accordingly. A scan of all possible boundary 
conditions, which can be parameterized in terms of a number of ({\it a priori} free) parameters, gives the following results: 
\begin{itemize}
\item Generically, that is if one does not tune any of the parameters, the resulting soft spectrum is that of gauge messengers mediation scenario. Gauginos acquire Dirac masses, while sfermions are loop suppressed, hence the soft spectrum turns out to be  similar to that of gaugino mediation models. These features arise due to the presence of massless (composite) bosonic and fermionic modes in the hidden sector.
\end{itemize}
There are two possible ways out of this scenario, whose effects we study in detail: 
\begin{itemize}
\item Tuning some of the parameters one can eliminate hidden sector
  massless modes: in this case one can scan any value of sfermion and
  gaugino masses, and obtain a spectrum which can range from gaugino
  mediation to minimal gauge mediation or scenarios with suppressed
  gaugino masses, hence covering all of GGM parameter space. 
\item A more constrained but attractive possibility, relies on adding dynamical degrees of freedom in the bulk. Within this approach, the hidden sector massless fermionic modes disappear automatically, and standard Majorana masses for gauginos are generated, without any necessary tuning of bulk field boundary conditions.
\end{itemize}

The basic objects one has to compute in GGM are two-point functions of  operators belonging to a hidden sector conserved current 
supermultiplet. For a $U(1)$ current these are parametrized in Euclidean signature and in momentum space as
\begin{align}
&\langle J(k)J(-k)\rangle=C_{0}(k^{2})\ , \label{GGM0}\\
&\langle j_{\alpha}(k)\bar{j}_{\dot\alpha}(-k)\rangle=-\sigma_{\alpha\dot\alpha}^{i}k_{i}C_{1/2}(k^{2})\ ,\label{GGM1/2}\\
&\langle j_{i}(k)j_{j}(-k)\rangle=(k_{i}k_{j}-\delta_{ij}k^{2})C_{1}(k^2)\ ,\label{GGM1}\\
&\langle j_{\alpha}(k)j_{\beta}(-k)\rangle=\epsilon_{\alpha\beta}B_{1/2}(k^{2})\ ,\label{GGMB}\
\end{align}
where a factor of $(2\pi)^4 \delta^{(4)}(0)$ is understood. Gaugino
and sfermion masses can be expressed in terms of the above correlation
functions as\footnote{Strictly speaking, eq.\eqref{Massgaugino} is 
approximate, the exact expression being given by the pole of the
resummed gaugino propagator. The exact value can differ
significantly from the approximate one when
$g^2 B_{1/2}$ and/or $g^2 C_{1/2}$ are ${\cal O}(1)$, as in some of the strongly coupled (large $N$) scenarios
that we consider.}
\begin{align}
&m_{\lambda} =g^2B_{1/2}(0)\label{Massgaugino}\ ,\\
&m^{2}_{\tilde{f}}=g^{2}q^{2}_{\tilde{f}}\int\frac{d^{4}k}{(2\pi)^4}\sum_{s=0,1/2,1}\frac{(-)^{2s}N_{s}}{k^2(1+g^2C_{s}(k^2))}\label{Masssfermion}\ ,
\end{align}
where $q_{\tilde{f}}$ is the sfermion $U(1)$ charge, $N_{0}=1$, $N_{1/2}=4$, $N_{1}=3$, and $g$ is the $U(1)$ gauge coupling. We have written the expressions (\ref{Massgaugino})--(\ref{Masssfermion}) following 
the spirit of \cite{Buican:2009vv, Intriligator:2010be}, in order to allow for massless poles in the $C_s$ correlators. We will indeed discuss cases where some or even 
all of them will have such a singularity. When all correlators are instead non-singular in the IR, one can expand the expression \eqref{Masssfermion} perturbatively in $g$, 
and obtain the usual leading order GGM expression for the sfermion masses derived in \cite{Meade:2008wd}. 

The program of holographic general gauge mediation outlined in \cite{Argurio:2012cd} consists in computing the correlators (\ref{GGM0})-(\ref{GGMB}) in strongly coupled hidden sectors using holography.


\subsection{Outline}

We begin in section 2 by 
recalling the techniques of holographic renormalization for AAdS spaces, when applied to a massless vector bulk supermultiplet. The latter is the holographic dual of  
the conserved current supermultiplets whose two-point functions we want to compute. This will allow us to provide an holographic expression for the two-point functions (\ref{GGM0})-(\ref{GGMB}), for a 
generic AAdS background. In order to set up notations, and as a useful reference for later studies, we 
will then apply such general formulas to the (superconformal) pure AdS case. In section 3 we move to consider the hard wall model, and show that one
can impose a variety of different boundary conditions on the bulk fields at the IR wall, obtaining a corresponding hidden sector in which the breaking of both supersymmetry and R-symmetry depends on independent parameters in the boundary conditions. The phenomenological
consequences of this freedom will be discussed in detail in section 4, where it will be shown when and how the general results we have anticipated above, can actually be achieved. A somewhat unpleasant feature of this way of proceeding is that in order to cover large portions of GGM parameter space, and in fact allow for non-vanishing Majorana gaugino masses to begin with, somewhat {\it ad hoc} boundary conditions should be imposed.  In section 5, by turning
on a background profile for a scalar whose dual operator is charged under the R-symmetry, we will show that gaugino masses can in fact be generated in a dynamical way, following a similar strategy as that of \cite{Argurio:2012cd}. We conclude in section 6 with a brief summary  and an outlook on further research directions. The appendices contain  a list of useful properties of Bessel functions that we use in our analysis, and the derivation of some analytic results about the behavior of GGM correlators at low momenta which we will use in the main text. 

\section{Holographic correlators for GGM}
Let us summarize the main requirements that a background has to satisfy in order to be interesting from our perspective. These requirements are directly related 
to specific features of the hidden sector gauge theory that we want to describe holographically: 
\begin{enumerate}
\item The requirement of Asymptotically Anti-de Sitter-ness results in having a gauge theory with a superconformal (interacting) UV fixed point. This is not a necessary 
physical requirement, but just a simplifying assumption we make to have more control on the AdS/CFT machinery, and which can be relaxed if one wants to consider other 
classes of (more realistic) hidden sectors.  
\item The background should asymptote AdS but deviate from it in the interior. This ensures the existence of one or more scales in the problem, which is a necessary condition for the background to break conformal invariance and supersymmetry. Single scale models are the simplest option, but more realistic gravitational duals might admit different (and unrelated) scales setting the breaking of conformal invariance and supersymmetry, respectively. Typically, the field theory will enjoy a confining behavior, with a mass gap.
\item The background has to support an $N=2$ vector multiplet
  corresponding to a gauged symmetry group - at least an
  $U(1)$ - in the bulk. This corresponds to an anomaly-free  global symmetry of the hidden sector, a necessary ingredient 
for a model of gauge mediation, where this global symmetry should eventually be weakly gauged. 
\end{enumerate}

The starting point for computing the GGM correlators (\ref{GGM0})-(\ref{GGMB}) holographically is to turn on the fluctuations for the fields dual to the conserved current supermultiplet, which, as just noticed,  form an $N=2$ vector multiplet in $\text{AdS}_{5}$. Since we are interested in two-point functions, we need the action for the fluctuating fields only to quadratic order.  In what follows, we will start considering a minimal action, with kinetic plus mass terms only
\begin{equation}
S_{\text{min}} = \frac{N^2}{4\pi^2}\int d^5x \sqrt{G}\left[ (G^{\mu\nu}\partial_{\mu}D\partial_{\nu}D-4D^2)+\frac{1}{4}F_{\mu\nu}F^{\mu\nu}+\frac{1}{2}(\bar{\lambda}\slashed{D}_{G}\lambda + c.c.) - \frac{1}{2}\bar{\lambda}\lambda\right]\ , \label{minimalaction}
\end{equation}
where we put the AdS radius to be $L=1$, and the masses $m_{A_\mu}=0$, $m^2_D=-4$ and $m_\lambda=-1/2$ are those required by the holographic dictionary. The equations of motion are 
\begin{align}
& (\Box_{G}+4)D=0 \ , \label{scalarminimal} \\
& \frac{1}{\sqrt{G}}\partial_{\mu}(\sqrt{G}G^{\mu\rho}G^{\nu\sigma}F_{\rho\sigma})=0\ , \label{vectorminimal}  \\
&(\slashed{D}_{G} - \frac{1}{2})\lambda=0 \ . \label{spinorminimal}
\end{align}
In a general AAdS background the asymptotics of the fluctuating fields in the vector multiplet is uniquely determined by their masses. In the coordinates 
\be
ds^2_5 = \frac{1}{z^2}\left( dz^2 + dx_i^2\right)\ , 
\ee
 the boundary is at $z\to0$,  and near the boundary the equations of motion become
\begin{align}
& (z^2\partial_{z}^2-3z\partial_{z}-z^2k^2+4)D=0 \ , \label{scalarminimalUV} \\
& (z^2\partial_{z}^2-z\partial_{z}-z^2k^2)A_{i}=0\ , \label{vectorminimalUV}  \\
&\begin{cases}
(z^{2}\partial_{z}^2-4z\partial_{z}-z^{2}k^2+\frac{21}{4})\bar{\xi}=0\ ,\\
z\bar{\sigma}^{i}k_{i}\chi=(z\partial_{z}-\frac{3}{2})\bar{\xi}\ .
\end{cases} \label{spinorminimalUV}
\end{align}
In order to obtain the equations above we have Fourier transformed the
fields, and we have fixed the gauge for the $5d$ gauge field taking $A_{z}=0$ and $-ik^{i}A_{i}=0$. Moreover, we have traded the first order equation of motion for a Dirac field with a second order equation of motion for one of its Weyl components plus a first order constraint for the other Weyl component, with 
\be
\lambda= \left(\begin{matrix}\chi \\ \bar \xi\end{matrix}\right)\ .
\ee
The resulting near boundary expansion is
\begin{align}
&\ \ \ D(z,k)\underset{\overset{{z\to0}}{}}{\simeq}z^2(\ln(\epsilon\Lambda)d_{0}(k)+\tilde{d}_{0}(k))+\mathcal{O}(z^4)\ , \label{boundary0}\\
&\ \ \ A_{i}(z,k)\underset{\overset{{z\to0}}{}}{\simeq}a_{i0}(k)+z^2(\tilde{a}_{i2}(k)+\frac{k^2}{2}\ln(z\Lambda)a_{i0}(k))+\mathcal{O}(z^4)\ ,\label{boundary1}\\
&\begin{cases}
\bar{\xi}(z,k)\underset{\overset{{z\to0}}{}}{\simeq}z^{3/2}(\bar{\xi}_{0}(k)+z^{2}(\bar{\tilde{\xi}}_{2}(k)+\frac{k^2}{2}\ln(z\Lambda)\bar{\xi}_{0}(k))+\mathcal{O}(z^4))\label{boundary1/2}\\
\chi(z,k)\underset{\overset{{z\to0}}{}}{\simeq}z^{5/2}(\tilde{\chi}_{1}-\ln(z\Lambda)\sigma^{i}k_{i}\bar{\xi}_{0}(k)+\mathcal{O}(z^{2}))    \ .
\end{cases}
\end{align}
The leading terms in the expansion $\lbrace d_{0}(k), a_{i0}(k), \bar{\xi}_{0}^{\dot{\alpha}}(k)\rbrace$ will be identified as the sources for the corresponding boundary operators $\lbrace J(k), j_{i}(k), \bar{j}_{\dot{\alpha}}(k)\rbrace$ and the undetermined subleading terms $\lbrace \tilde{d}_{0}(k), \tilde{a}_{i2}(k), \tilde{\chi}_{1\alpha}(k)\rbrace$ will be related to the one-point functions  of the same boundary operators. 
The near boundary limit of the action \eqref{minimalaction} appropriately renormalized gives us the $4d$ action from which one can derive the holographic expression for the GGM two point functions 
\begin{align} 
&\langle J(k)J(-k)\rangle=\frac{N^2}{8\pi^2}\left(-  2\frac{\delta \tilde{d}_0}{\delta d_0} \right)                \  , \label{Ch0} \\
&\langle j_{i}(k)j_{j}(-k)\rangle=\frac{N^2}{8\pi^2} \left(2\frac{\delta \tilde{a}_{i2}}{\delta{a_0^j}} + 2\frac{\delta \tilde{a}_{j2}}{\delta{a_0^i}} +   k^2 \delta_{ij} \right)      \  , \label{Ch1} \\
&\langle j_{\alpha}(k)\bar{j}_{\dot{\alpha}}(-k)\rangle=\frac{N^2}{8\pi^2}\left(  \frac{\delta\tilde{\chi}_{1\alpha}}{\delta\bar{\xi}^{\dot{\alpha}}_0} +  \frac{\delta\bar{\tilde{\chi}}_{1\dot\alpha}}{\delta\xi^{\alpha}_0}  \right) =\frac{N^2}{8\pi^2}\left[ \frac{(\sigma^i k_i)_{\alpha\dot{\beta}}}{k^2}\left(2\frac{\delta\bar{\tilde{\xi}}_2^{\dot{\beta}}}{\delta\bar{\xi}_0^{\dot{\alpha}}} + \frac{k^2}{2}\delta^{\dot{\beta}}_{\dot{\alpha}}\right) + c.c. \right]    \  , \label{Ch1/2}\\
&\langle j_{\alpha}(k)j_{\beta}(-k)\rangle =\frac{N^2}{8\pi^2}\left(\frac{\delta\tilde{\chi}_{1\alpha}}{\delta\xi_{0}^{\beta}} - \frac{\delta\tilde{\chi}_{1\beta}}{\delta\xi_{0}^{\alpha}}\right)=\frac{N^2}{8\pi^2}\left(2\frac{(\sigma^i k_i)_{\alpha\dot{\alpha}}}{k^2} \frac{\delta \bar{\tilde{\xi}}_2^{\dot{\alpha}}}{\delta \xi_0^{\beta}} - (\alpha \leftrightarrow \beta) \right). \label{Bh}
\end{align}
The fermionic correlators \eqref{Ch1/2} and \eqref{Bh} are naturally expressed in terms of the subleading mode $\tilde{\chi}_{1}(k)$ of the positive chirality field $\chi(z,k)$. For later convenience, we have also re-expressed the formulas in terms of the subleading mode $\tilde{\xi}_{2}(k)$ of the negative chirality field, which is related to $\tilde{\chi}_{1}(k)$ by the equations of motion.

\subsection{Pure AdS: the superconformal reference results}
As a reference for what we do next, let us now review the standard supersymmetric AdS computation of \cite{Argurio:2012cd}. 
The equations of motion for the fluctuating fields in $\text{AdS}_{5}$ are exactly the second order linear ODE (\ref{scalarminimalUV})--(\ref{spinorminimalUV}) defined in the open set $(0,\infty)$. Performing a rescaling of the bulk fields we can bring these equations to the standard form for the Bessel equation (see appendix \ref{A} for a review of the essential features of Bessel functions). In fact, taking 
\begin{equation}
D=z^{2}d\ ,\ A_{i}=z\alpha_{i}\ ,\ \bar{\xi}=z^{5/2}\bar{\Xi}\ ,\ \chi=z^{5/2}X\ , \label{soleqdiff}
\end{equation}  
we get for the equations of motion
\begin{align}
&(z^2\partial_{z}^2+z\partial_{z}-z^2k^2)d=0 \ , \\
&(z^2\partial_{z}^2+z\partial_{z}-(z^2k^2+1))\alpha_{i}=0\ ,  \\
&\begin{cases}
(z^{2}\partial_{z}^2+z\partial_{z}-(z^{2}k^2+1))\bar{\Xi}=0\ ,\\
z\bar{\sigma}^{i}k_{i}X=(z\partial_{z}+1)\bar{\Xi}\ ,
\end{cases} 
\end{align}
whose general solutions for $k^2\ge0$ are  
\begin{align}
&d(z,k)=c_{1}(k)I_{0}(kz)+c_{2}(k)K_{0}(kz)\ , \label{general0}\\
&\alpha_{i}(z,k)=\alpha_{i1}(k)I_{1}(kz)+\alpha_{i2}(k)K_{1}(kz)\ , \label{general1}\\
&\bar{\Xi}(z,k)=\bar{\theta}_{1}(k)I_{1}(kz)+\bar{\theta}_{2}(k)K_{1}(kz)\ .\label{general1/2}
\end{align}
The general solutions of second order differential equations depend on two arbitrary constants, that in our case can be arbitrary functions of the $4d$ momentum $k$. 

In order to impose Dirichlet boundary conditions at  $z\to0$, which identify the sources for the boundary operators, we expand the solutions (\ref{general0})--(\ref{general1/2}) using (\ref{Ismall})--(\ref{Ksmall}).  Comparing with (\ref{boundary0})--(\ref{boundary1/2}) we get
\begin{equation}\label{UVbc}
c_{2}(k)=-d_{0}(k)\ ,\ \alpha_{i2}=ka_{i0}(k)\ ,\ \bar{\theta_{2}}=k\bar{\xi}_{0}(k)\ .
\end{equation}
This leaves us with three arbitrary functions $c_{1}(k), a_{i1}(k), \theta_{1}(k)$. In order to have a regular solution in the full domain we have to impose the following conditions in the deep interior
\begin{equation}
\lim_{z\to\infty}D(z,k)=0\ ,\ \lim_{z\to\infty}A_{i}(z,k)=0\ ,\ \lim_{z\to\infty}\xi(z,k)=0\ . \label{regularityAdS}
\end{equation}
Using the expansions (\ref{largeI})--(\ref{largeK}) it is easy to see that \eqref{regularityAdS} implies $c_{1}(k)=a_{i1}(k)=\theta_{1}(k)=0$. 
We note that the bulk boundary conditions are crucial to single out the solution to the fluctuation equations. The pure $\text{AdS}_{5}$ solutions are
\begin{align}
&D^{AdS}(z,k)=-z^{2}K_{0}(kz)d_{0}(k)\ ,\\
&A^{AdS}_{i}(z,k)=zkK_{1}(kz)a_{i0}(k)\ ,\\
&\bar{\xi}^{AdS}(z,k)=z^{5/2}kK_{1}(kz)\bar{\xi}_{0}(k)\ ,\\
&\bar{\chi}^{AdS}(z,k)=z^{5/2}K_{0}(kz)\sigma^{i}k_{i}\bar{\xi}_{0}(k)\ , 
\end{align}
where in order to derive the last equation we have used the recurrence relations among the Bessel functions  \eqref{recurrence}. 

Expanding these results near the boundary we get the usual $\text{AdS}_{5}$ result for the subleading modes from which we derive the GGM two-point functions 
\begin{align}
&\langle J(k)J(-k)\rangle=\frac{N^2}{8\pi^2} C_{0}^{AdS}(k^2)=\frac{N^2}{8\pi^2} \left[\ln\left(\frac{\Lambda^2}{k^2}\right)+2\ln2-2\gamma\right]\ , \label{AdSC0}\\
&\langle j_{i}(k)j_{j}(-k)\rangle=-\frac{N^2}{8\pi^2}\delta_{ij}k^2C_{1}^{AdS}(k^2)=-\frac{N^2}{8\pi^2} \delta_{ij}k^2\left[\ln\left(\frac{\Lambda^2}{k^2}\right)+2\ln2-2\gamma\right]\ , \label{AdSC}\\
&\langle j_{\alpha}(k)\bar{j}_{\dot{\alpha}}(-k)\rangle=-\frac{N^2}{8\pi^2}\sigma^{i}k_{i}C_{1/2}^{AdS}(k^2)=-\frac{N^2}{8\pi^2} \sigma^{i}k_{i}\left[\ln\left(\frac{\Lambda^2}{k^2}\right)+2\ln2-2\gamma\right]\ ,\\
&\langle j_{\alpha}(k)j_{\beta}(-k)\rangle=0\ , \label{AdSB}
\end{align}
where we can always take in \eqref{AdSC} $\delta_{ij}\rightarrow\delta_{ij}-\frac{k_{i}k_{j}}{k^2}$ because of current conservation.\footnote{As was already noticed in \cite{Argurio:2012cd} the holographic prescription performed with our gauge fixing allows us to compute only the part of the vector current two-point function which is proportional to $\delta_{ij}.$}

As expected in a supersymmetric background, $B_{1/2}=0$ and $C_{0}^{AdS}=C_{1}^{AdS}=C_{1/2}^{AdS}=C^{AdS}$ so that both the Majorana gaugino mass \eqref{Massgaugino} and the sfermion masses \eqref{Masssfermion} are identically zero. It is worth noticing that the correlators (\ref{AdSC0})-(\ref{AdSB}) reflect the fact that the hidden sector is exactly superconformal, exhibiting a branch-cut at $k=0$ which corresponds to the two-particle exchange of the $2N^2$ massless particles of the hidden sector which are charged under the gauged $U(1)$.

\section{Hard wall models}

Let us now consider the model that will be our primary interest in this paper, the hard wall model. 
This is just $\text{AdS}_{5}$ in which the geometry ends abruptly in 
the interior by putting a sharp IR cut-off at $z=1/\mu$. This model
was originally studied as a toy model of a confining gauge theory
because it provides an holographic dual for theories with a gapped and
discrete spectrum \cite{Polchinski:2001tt, BoschiFilho:2002vd,
  Erlich:2005qh}. Our aim is to study the behavior of GGM two-point
functions on this background.\footnote{Let us notice that our set up is reminiscent of extra dimensional scenarios like \cite{Gherghetta:2000kr,McGarrie:2010yk,Abel:2010vb} in which, however, the physics of the $4d$ hidden sector arises as a KK reduction of a $5d$ theory in a slice of AdS.} Not surprisingly, the behavior of the correlators will depend strongly on the boundary conditions that one has to  impose on the bulk fields at the IR cut-off.

In the case of an hard wall background the general solution of the equations of motion for the fluctuations in the vector multiplet is exactly the same as for pure AdS, eqs.  (\ref{general0})--(\ref{general1/2}), and depends on six integration constants (two for each field). Also the UV boundary conditions (\ref{UVbc}) remain the same, and can be simply understood as fixing the source of the boundary operator, leaving only three constants undetermined.  
However, we are now solving the differential equations in the open set
$(0,1/\mu)$, and the regularity conditions are replaced by some IR
boundary conditions at $z=1/\mu$. These conditions can be solved for
the three remaining constants, in order to fix the functional dependence of the subleading modes on the leading ones.

The generic form of the solutions is
\begin{align}
& D(z,k) = D^{AdS}(z,k) + c_{1}(k)z^2I_{0}(kz) \ , \label{HWD}\\
& A_{i}(z,k) = A_{i}^{AdS}(z,k) + \alpha_{1i}(k) z I_{1}(kz) \ , \label{HWA}\\
& \bar{\xi}(z,k) = \bar{\xi}^{AdS}(z,k) + \bar{\theta}_{1}(k) z ^{5/2} I_{1}(kz) \ , \label{HWxi}
\end{align}
where $c_{1}$, $\alpha_{i1}$ and $\bar{\theta}_1$ are integration constants determined by the IR boundary conditions. Expanding these expressions near the UV boundary, one can easily find that the correlators are modified with respect to the AdS case in the following way
\begin{align}
&C_{0}(k^2) = C^{AdS}(k^2) -2\frac{\delta c_{1}}{\delta{{d}_0}} \ , \label{HWC0}\\
&C_{1}(k^2) = C^{AdS}(k^2) - \frac{1}{2k}\delta ^i _j  \frac{\delta \alpha _{1i}}{\delta a_{0j}} \ , \\
&C_{1/2}(k^2) = C^{AdS}(k^2) - \left( \frac{1}{2k}\delta ^{\dot{\alpha}} _{\dot{\beta}}  \frac{\delta \bar{\theta}_{1}^{\dot{\beta}}}{\delta \bar{\xi}_0^{\dot{\alpha}}}+ c.c. \right) \ , \\
& B_{1/2}(k^{2}) = - \frac{\sigma ^i_{\alpha\dot{\alpha}}k_i}{k}\frac{\delta \bar{\theta}_{1}^{\dot{\alpha}}}{\delta \xi _{0 \alpha}}. \label{HWB}
\end{align}

\subsection{Homogeneous IR boundary conditions}
We start by taking general homogeneous boundary conditions at the IR cut-off
\begin{align}
&(D(z,k)+\rho_{0}z\partial_{z}D(z,k))\vert_{z=1/\mu}=0\ , \label{hwIR0}\\
&(A_{i}(z,k)+\rho_{1}z\partial_{z}A_{i}(z,k))\vert_{z=1/\mu}=0\ , \label{hwIR1}\\
&(\bar{\xi}(z,k)+\rho_{1/2}z\partial_{z}\bar{\xi}(z,k))\vert_{z=1/\mu}=0\ ,\label{hwIR1/2}
\end{align} 
which depend on three independent coefficients $\rho_{s}$. As we will see, in order to cover all of GGM parameter space it will be necessary to turn on also inhomogeneous terms in the above equations, something we will do next.   

As it befits coefficients computed with homogeneous boundary conditions, the coefficients $c_1$, $\alpha_{1i}$ and $\bar{\theta}_1$ in (\ref{HWD})--(\ref{HWxi}) are all proportional to the source terms. The resulting GGM functions are
\begin{align}
&C_{0}^{(h)}(k^2) = C^{AdS}(k^2) + 2 \frac{-(1+2\rho_{0})K_{0}(\frac{k}{\mu})+\rho_{0}\frac{k}{\mu} K_{1}(\frac{k}{\mu})}{(1+2\rho_{0})I_{0}(\frac{k}{\mu})+\rho_{0}\frac{k}{\mu}I_{1}(\frac{k}{\mu})}\ , \label{1Hard0}\\
&C_{1}^{(h)}(k^2) = C^{AdS}(k^2)+2\frac{K_{1}(\frac{k}{\mu})-\rho_{1}\frac{k}{\mu}K_{0}(\frac{k}{\mu})}{I_{1}(\frac{k}{\mu})+\rho_{1}\frac{k}{\mu}I_{0}(\frac{k}{\mu})}\ ,\label{1Hard1}\\
&C_{1/2}^{(h)}(k^2) = C^{AdS}(k^2)+2\frac{(1+\frac{3}{2} \rho_{1/2})K_{1}(\frac{k}{\mu})- \rho_{1/2}\frac{k}{\mu}K_{0}(\frac{k}{\mu})}{(1+\frac{3}{2} \rho_{1/2})I_{1}(\frac{k}{\mu})+\rho_{1/2}\frac{k}{\mu}I_{0}(\frac{k}{\mu})}\ , \label{1Hard1/2}\\ 
&B_{1/2}^{(h)}(k^2)=0\ . \label{1HardB}
\end{align}

The analysis of the boundary condition-dependent soft spectrum emerging from the correlators (\ref{1Hard0})--(\ref{1Hard1/2}) is postponed to section \ref{phenohw}. For future reference we would instead like to comment here on the behavior of the correlators in the IR and UV. Making use of the asymptotic expansion for $x\ll 1$ for the Bessel functions (\ref{Ismall})--(\ref{Ksmall}) we find the correlators at low momentum to behave as
\begin{align}
&C_{0}^{(h)}(k^2)\underset{\overset{{k\to0}}{}}\simeq \ln\left(\frac{\Lambda^2}{\mu^2}\right)+\frac{2\rho_{0}}{1+2\rho_{0}} \label{hIR0}\ ,\\
&C_{1}^{(h)}(k^2)\underset{\overset{{k\to0}}{}}\simeq \frac{4}{1+2\rho_{1}}\frac{\mu^2}{k^{2}}+\ln\left(\frac{\Lambda^2}{\mu^2}\right)-\frac{3+8\rho_{1}}{2(1+2\rho_{1})^2}\label{hIR1}\ ,\\ 
&C_{1/2}^{(h)}(k^2)\underset{\overset{{k\to0}}{}}\simeq 4\frac{2+3\rho_{1/2}}{2+7\rho_{1/2}}\frac{\mu^2}{k^2}+\ln\left(\frac{\Lambda^2}{\mu^2}\right)-\frac{(2+3\rho_{1/2})(6+25\rho_{1/2})}{2(2+7\rho_{1/2})^2}\ . \label{hIR1/2}
\end{align}
As for the UV limit, given the large $x$ behavior of Bessel functions \eqref{largeI}--\eqref{largeK}, we can see that all the $C_s^{(h)}$  functions approach the supersymmetric AdS value with exponential rate at large momentum
\be
C_0^{(h)}(k^2)\sim C_{1/2}^{(h)}(k^2)\sim C_1^{(h)}(k^2)\underset{\overset{{k\to\infty}}{}}\simeq C^{AdS}(k^2)-2\pi e^{-\sqrt{\frac{k^2}{\mu^2}}} \ .
\ee
From the field theory point of view, the exponential suppression of the SUSY-breaking effects in the UV suggests that SUSY-breaking in a hidden sector described by a hard wall holographic model is not induced by any quantum operator, which generically would appear in the OPE of the GGM correlation functions with a scaling behavior in $k^2$ fixed by its dimension.  

Two additional remarks are in order at this point. The first is that one can of course compute the above functions also using the numerical approach pursued in \cite{Argurio:2012cd}, finding perfect agreement with the analytic computation above. The second comment is that the above functions can be continued to negative values of $k^2$. It is easy to convince oneself that they will then display an infinite sequence of poles on the negative $k^2$ axis, corresponding to the glueball towers for each spin sector. They return the same values that can be obtained through the more traditional holographic approach of computing glueball masses, i.e. finding normalizable  fluctuations for each field.

\subsection{Inhomogeneous IR boundary conditions}
Let us now consider the possibility of having inhomogeneous boundary conditions in the IR.
We thus take general boundary conditions at the IR cut-off depending
on three more arbitrary terms, now
\begin{align}
&(D(z,k)+\rho_{0}z\partial_{z}D(z,k))\vert_{z=1/\mu}=\Sigma_{0}(k)\ , \label{hwIR0nh}\\
&(A_{i}(z,k)+\rho_{1}z\partial_{z}A_{i}(z,k))\vert_{z=1/\mu}=\Sigma_{i1}(k)\ , \label{hwIR1nh}\\
&(\bar{\xi}(z,k)+\rho_{1/2}z\partial_{z}\bar{\xi}(z,k))\vert_{z=1/\mu}=\bar{\Sigma}_{1/2}(k)\  ,\label{hwIR1/2nh}
\end{align} 
where we have allowed for a non trivial $k$ dependence in the
inhomogeneous terms $\Sigma_{s}$. We will see instantly that the
arbitrariness actually amounts to 4 new constants. 

The coefficients $c_{1}$, $\alpha_{i1}$ and $\bar{\theta}_1$ in eqs. (\ref{HWD})--(\ref{HWxi}) will pick up an additional contribution, linear in the $\Sigma_{s}$. Since these coefficients enter the GGM functions only through the first derivative with respect to the source, the inhomogeneous terms can contribute only if we allow them to be dependent on the source, with the result that the condition at $z = 1/\mu$ involves both IR and UV data of the function. In particular, from eq. (\ref{HWB}), a dependence of $\bar\Sigma_{1/2}(k)$ on the source $\xi_{0}$ can give a non-vanishing $B_{1/2}$,  as opposed to the case of homogenous boundary conditions (\ref{1HardB}). Therefore, such a dependence implies that the boundary condition \eqref{hwIR1/2nh} explicitly breaks the R-symmetry.

Since in any case only the first derivative enters eqs.(\ref{HWC0})-(\ref{HWB}), it is enough to
let the $\Sigma_{s}$ depend linearly on the sources $d_0(k),
a_{i0}(k)$ and $\xi_0(k)$. Taking into account Lorentz invariance, a reasonable choice is
\begin{align}
&\Sigma_{0}(k^2)=-\frac{1}{\mu^2}E_{0}d_{0}(k)\ ,\\
&\Sigma_{i1}(k^2)=-E_{1}a_{i0}(k)\ ,\\
&\bar{\Sigma}_{1/2}^{\dot\alpha}(k^2)=-\frac{1}{\mu^{3/2}}E_{1/2}\bar{\xi}^{\dot\alpha}_{0}(k)-H_{1/2}\frac{1}{\mu^{7/2}}\bar{\sigma}_{i}^{\dot{\alpha}\alpha}k^{i}\xi_{\alpha0}(k)\ , \label{pippa}
\end{align}
where the $E$'s and $H$ are coefficients which do not depend on the
momentum. Hence we are left with 4 new
parameters due to the inhomogeneous boundary conditions.

The GGM functions in this case take the form
\begin{align}
&C_{0}^{(nh)}(k^2) = C^{AdS}(k^2) + 2 \frac{-(1+2\rho_{0})K_{0}(\frac{k}{\mu})+\rho_{0}\frac{k}{\mu} K_{1}(\frac{k}{\mu})+E_0}{(1+2\rho_{0})I_{0}(\frac{k}{\mu})+\rho_{0}\frac{k}{\mu}I_{1}(\frac{k}{\mu})}\ , \label{1Hard0nh}\\
&C_{1}^{(nh)}(k^2) = C^{AdS}(k^2)+2\frac{K_{1}(\frac{k}{\mu})-\rho_{1}\frac{k}{\mu}K_{0}(\frac{k}{\mu})+\frac{\mu}{k} E_1}{I_{1}(\frac{k}{\mu})+\rho_{1}\frac{k}{\mu}I_{0}(\frac{k}{\mu})}\ ,\label{1Hard1nh}\\
&C_{1/2}^{(nh)}(k^2) = C^{AdS}(k^2)+2\frac{(1+\frac{3}{2} \rho_{1/2})K_{1}(\frac{k}{\mu})- \rho_{1/2}\frac{k}{\mu}K_{0}(\frac{k}{\mu})+\frac{\mu}{k} E_{1/2}}{(1+\frac{3}{2} \rho_{1/2})I_{1}(\frac{k}{\mu})+\rho_{1/2}\frac{k}{\mu}I_{0}(\frac{k}{\mu})}\ , \label{1Hard1/2nh}\\ 
&B_{1/2}^{(nh)}(k^2)=2\frac{\frac{k}{\mu} H_{1/2}}{(1+\frac{3}{2} \rho_{1/2})I_{1}(\frac{k}{\mu})+\rho_{1/2}\frac{k}{\mu}I_{0}(\frac{k}{\mu})}\ . \label{1HardBnh}
\end{align}
The result with homogeneous boundary condition is simply recovered by setting the $E$'s and $H$ to zero.

The inhomogeneous terms contribute to the IR behavior as follows
\begin{align}
&C_{0}^{(nh)}(k^2)-C_{0}^{(h)}(k^2)\underset{\overset{{k\to0}}{}}\simeq\frac{2}{1+2\rho_{0}}E_0\ ,\label{inIR0}\\
&C_{1}^{(nh)}(k^2)-C_{1}^{(h)}(k^2)\underset{\overset{{k\to0}}{}}\simeq\frac{4}{1+2\rho_{1}}\frac{\mu^{2}}{k^{2}}E_1\label{inIR1}\ ,\\
&C_{1/2}^{(nh)}(k^2)-C_{1/2}^{(h)}(k^2)\underset{\overset{{k\to0}}{}}\simeq\frac{8}{2+7\rho_{1/2}}\frac{\mu^{2}}{k^{2}}E_{1/2}\label{inIR1/2}\ ,\\
&B_{1/2}^{(nh)}(k^2)\underset{\overset{{k\to0}}{}}\simeq\frac{8}{2+7\rho_{1/2}}H_{1/2}\ .
\end{align}
In particular, having $H_{1/2}\neq0$ we get now a non-zero Majorana
mass for the gaugino. Indeed, the boundary condition (\ref{pippa})
explicitly breaks the R-symmetry. 

As for the UV asymptotic, the large $x$ behavior of the Bessel functions \eqref{largeI} tells us that the exponential approach to the supersymmetric limit remains valid in this case, also for $B_{1/2}(k^2)$ that asymptotes to 0. So we see that, consistently, the inhomogeneous boundary conditions do not modify the UV behavior.

\section{Analysis of the soft spectrum}
\label{phenohw}

We now discuss the physical interpretation, in terms of soft supersymmetry breaking masses, of the $C_s$ and $B$ functions we have found in the previous section. 

Let us start with a very basic requirement: since the correlators happen to have non-trivial denominators which depend on the momentum, we should exclude the possibility that tachyonic poles are developed. The denominators are linear combinations of two Bessel functions evaluated at $x=k/\mu$, and studying their monotonicity properties and their limits for $x \to 0$ and $x \to \infty$ one can easily see that the poles are excluded if and only if the coefficients of the linear combination have the same sign. This condition results in the following inequalities
\be\label{tac}
 \{ \rho_0 \leq -\frac12 \} \cup \{ \rho_0 \geq 0\} \ ,  \{\rho_1 \geq 0\} \ , \{ \rho_{1/2} \leq - \frac 23\} \cup \{\rho_{1/2} \geq 0\} \ . 
\ee

The IR behavior of the $C_s$ functions, in particular the expressions given in eqs. \eqref{hIR0}--\eqref{hIR1/2}, shows that the theory described holographically by the hard wall  has a threshold $\mu$ for the production of two particle states and possibly a certain number of massless poles which depends on the choice of the boundary conditions. Below we analyze the cases of homogeneous and inhomogeneous boundary conditions in turn.

\subsection{Homogeneous boundary conditions}

For generic choices of $\rho_s$ parameters, we see from eqs. \eqref{hIR0}--\eqref{hIR1/2} that $C_1$ and $C_{1/2}$ have poles at $k^2 = 0$ while $C_0$ has not. The interpretation of such poles is that they arise from the exchange of a massless state with the same quantum numbers of the corresponding operator. In $C_1$, this means that the global $U(1)$ symmetry is spontaneously broken, the massless excitation being the associated Goldstone boson. If the symmetry is broken, we cannot identify it with the Standard Model gauge group, but rather with an extension thereof by some higgsed $U(1)^{\prime}$, a setting extensively studied in the literature (see for instance \cite{Langacker:2008yv} and references therein).

The pole in $C_{1/2}$ signals the existence of an R-charged massless fermion, neutral under the global $U(1)$, which mixes with the fermionic partner of the current. The most natural interpretation of such a fermion in a strongly coupled theory is that of  a 't Hooft fermion associated with a global anomaly of the unbroken $U(1)_R$.

The consequence on the soft spectrum of poles in the correlators $C_1$ and/or $C_{1/2}$ was studied in \cite{Buican:2009vv,Intriligator:2010be}, and can be summarized as follows: the gaugino acquires a Dirac mass by mixing with the would-be massless fermion in $C_{1/2}$ (recall that a Majorana mass is forbidden by the unbroken R-symmetry), and the integral giving the sfermion masses is dominated by the contribution of the poles. Comparing with the usual result in General Gauge Mediation without IR singularities, the sfermion soft mass is enhanced  by a logarithm of the gauge coupling. Notice that the pole in $C_1$  ($C_{1/2}$) contributes with a negative (positive) sign, so that generically one can get a tachyonic contribution to the sfermion mass-squared. In formulae
\begin{align}
& m_{\tilde g} = g M_{1/2} \  , \label{Dirgmass}\\
& m_{\tilde f}^2  \simeq  \frac{g^4}{(4\pi)^2}\left(\log \frac{1}{g^2}\right)(4 M_{1/2}^2 - 3 M_1^2)  \ , \label{sfmass}  
\end{align}
where $g$ is the gauge coupling, $m_{\tilde g}$ is the Dirac mass of the gaugino, $m_{\tilde f}$ is the sfermion mass, and $M_s^2$ is the residue of the massless pole $C_s \simeq M_s ^2 / k^2$. From eqs. (\ref{hIR1})--(\ref{hIR1/2}) we see that in our model\footnote{Here and in the following we tacitly assume that the prefactor $N^2/8\pi^2$ can be set to unity. For this value the overall normalization of the physical correlation functions \eqref{GGM0}--\eqref{GGMB} coincides with the one that we have used throughout \eqref{Ch0}--\eqref{Bh}. }
\be
M^2_1 = 4\mu^2\frac{1}{1+2\rho_1} \ , \ M^2_{1/2} = 4\mu^2 \frac{1+ \frac 32 \rho_{1/2}}{1+ \frac 72 \rho_{1/2}}. 
\ee
Notice that in the tachyon-free range (\ref{tac}) the two residues are
always positive. If we further impose the contribution to the sfermion mass-squared (\ref{sfmass}) to be positive, we get the additional inequality
\be
\rho_1 \geq -\frac 18 \  \frac{1 - \frac 92 \rho_{1/2}}{1+ \frac 32 \rho_{1/2} }.
\ee
We see from eqs. (\ref{Dirgmass})--(\ref{sfmass}) that in this scenario the sfermions are somewhat lighter that the gaugino. This is typical of Dirac gaugino scenarios \cite{Benakli:2008pg}, though in our model the Dirac partner of the gaugino is a strongly coupled composite fermion.

\subsubsection*{Tuning the $\rho_s$ parameters}

We now briefly mention different possibilities to evade the generic scenario presented above, which can be realized by choosing specific values for the $\rho_s$ parameters.

\begin{enumerate}
\item  As a first possibility, consider the case in which $M_1^2 = M_{1/2}^2$, that is
\be\label{eqres}
\rho_1 = \frac{\rho_{1/2}}{1 + \frac 32 \rho_{1/2}}\ ,
\ee
while $\rho_0$ is kept generic. We are still in a scenario in which the global symmetry is spontaneously broken in the hidden sector, and the soft spectrum is described by the same formulae as before (notice however that the contribution to the sfermion mass-squared is positive, now). Nevertheless, in this case we can argue a different interpretation of the physics in the hidden sector, the reason stemming from a somehow surprising fact: the condition (\ref{eqres}) that makes the two residues coincide, actually renders the whole $C_1$ and $C_{1/2}$ functions (\ref{1Hard1}) and (\ref{1Hard1/2}) equal for all values of $k^2$. As a consequence, one is led to interpret the massless fermion as the partner of the Goldstone boson associated to the broken global symmetry, rather than a 't Hooft fermion. Since $C_0$ differs from $C_1=C_{1/2}$ for generic $\rho_0$, supersymmetry is still broken in the hidden sector, but mildly enough so not to lift the fermionic partner of the Goldstone boson.  
\item As a subcase of 1, consider in addition to tune the $\rho_0$ parameter to $\rho_0 = -1/2$. In this case the low momentum expansion (\ref{hIR0}) is not valid, and by repeating the analysis one finds that also $C_0$ develops a $1/k^2$ pole, with residue $M_0^2 = 4\mu^2$. As explained in \cite{Buican:2009vv,Intriligator:2010be}, a pole in $C_0$ is unphysical, unless the hidden sector breaks the global symmetry in a supersymmetric manner, so that $C_0 = C_{1/2} = C_1$ and a massless Goldstone mode is present in all three functions\footnote{In the simple example of a $U(1)$ broken by the VEV of a charged chiral superfield the pole in $C_0$ is related to the modulus of the complex scalar.}.  Indeed, if we require $M_0^2=M_{1/2}^2=M_1^2$, that is $\rho_1 = \rho_{1/2} = 0$, we find from eqs. (\ref{1Hard0})--(\ref{1Hard1/2}) that this condition is sufficient to ensure $C_0 = C_{1/2}= C_1$ for all values of $k^2$, supporting the interpretation of a supersymmetric global symmetry breaking in the hidden sector.
\item Finally, $\rho_1$ and $\rho_{1/2}$ can also be (independently) tuned in such a way to eliminate the massless pole in $C_1$ and $C_{1/2}$ respectively, the specific values being $\rho_1 = \infty$\footnote{A global parametrization which avoids infinities could be conveniently given in terms of angles $\alpha_s$, the change of variable being $\rho_s = \mathrm{tg}( \alpha _s)$.} and $\rho_{1/2}= -2/3$. If only one of the two parameters is tuned, the soft masses and the interpretation of the physics in the hidden sector remains the same as in the previous section, with the only difference that $M_1^2$ or $M_{1/2}^2$ are tuned to $0$. It is therefore more interesting to consider the possibility that both parameters are tuned: in this case none of the $C_s$ has an IR singularity and we are in a situation similar to ordinary GGM, as far as sfermion masses are concerned  (the gaugino remains massless because the hidden sector does not break the R-symmetry). Since at large $k$ all the $C_{s}$ approach their supersymmetric value exponentially, the weighted sum $-(C_0 - 4 C_{1/2} + 3 C_{1/2})$ goes to zero at the same rate, so that we can determine the sign of the sfermion mass-squared by studying its IR limit. From eqs. (\ref{hIR0})--(\ref{hIR1/2}) we see that the leading term, with the present values of $\rho_1$ and $\rho_{1/2}$, is given by
\be\label{weisumIR}
-(C_0 - 4 C_{1/2} + 3 C_{1}) \underset{\overset{{k\to0}}{}}\simeq - \frac{2 \rho_0}{1 + 2 \rho_0} \ .
\ee
In the tachyon-free range (\ref{tac}) this expression is
negative. Therefore, in this tuned scenario we find vanishing gaugino
mass and tachyonic sfermion mass. We will see later  that both this
unwanted features can be overcome: one way, which is somehow more {\it ad-hoc}, consists in enlarging the parameter space by considering inhomogeneous boundary conditions; the other, which is more dynamical, consists in turning on a R-breaking scalar on top of the hard wall background. Most of what follows will therefore consist in improvements of this setting with tuned $\rho_1$ and $\rho_{1/2}$.
\end{enumerate}

\subsection{Inhomogeneous boundary conditions}

Let us proceed by considering the functions (\ref{1Hard0nh})--(\ref{1Hard1/2nh}), which we obtained adding source-dependent inhomogeneous terms in the boundary condition. Besides the $\rho_s$, we have now four additional real parameters to play with, namely the dimensionless $E_s$ and the R-breaking parameter $H_{1/2}$, which has dimension of a mass.

For generic values of the parameters the situation is analogous to the one with homogeneous boundary conditions, so that the $E_s$ parameters appear to be somehow redundant: $C_1$ and $C_{1/2}$ have a massless pole, while $C_0$ has not. The major difference with respect to the previously considered case is that now $H_{1/2}$ gives a non-zero Majorana mass to the gaugino,
\be
m_{\tilde g} = \frac{8}{2+7\rho_{1/2}}H_{1/2} \ .
\ee
Since now R-symmetry is broken, the pole in $C_{1/2}$ cannot be interpreted as due to a 't\,Hooft fermion, and it seems unphysical. 
In order to get more interesting and reasonable results, eliminating the poles  at $k^2=0$ in $C_1^{(nh)}$ and $C_{1/2}^{(nh)}$, we can take $E_1=-1$ and $E_{1/2}=-(1+\frac{3}{2} \rho_{1/2})$, see eqs. (\ref{inIR1}) and (\ref{inIR1/2}). As opposed to eq. (\ref{weisumIR}), the IR limit of the weighted sum $-(C_0 - 4 C_{1/2} + 3 C_{1/2})$ depends now on four parameters, the $\rho_s$ and $E_0$, so that one can easily obtain a positive mass-squared for the sfermions. For definiteness and for an easier comparison with eq. (\ref{weisumIR}), consider taking $\rho_1=\infty$ and $\rho_{1/2}=-2/3$, so that
\be
-(C_0 - 4 C_{1/2} + 3 C_{1}) \underset{\overset{{k\to0}}{}}\simeq - \frac{2 \rho_0 + 2E_0}{1 + 2 \rho_0} \ ,\label{erho}
\ee 
which can be positive if $E_0<-\rho_0$ (assuming a positive $\rho_0$). The sfermion masses can then be 
even bigger than the Majorana gaugino mass if $H_{1/2}$ is somewhat smaller than $\sqrt{|E_0|}\mu$. 

The punchline of the above analysis is that tuning appropriately the boundary conditions, one can realize holographically any scenario between pure gaugino mediation \cite{Kaplan:1999ac, Chacko:1999mi, Csaki:2001em, Cheng:2001an, Green:2010ww, McGarrie:2010qr, Sudano:2010vt, Auzzi:2010mb} to minimal gauge mediation \cite{Dvali:1996cu, Dimopoulos:1996gy, Martin:1996zb} as well as scenarios with suppressed gaugino masses \cite{Randall:1996zi, Csaki:1997if, ArkaniHamed:1998kj, Seiberg:2008qj, Elvang:2009gk, Argurio:2009ge, Argurio:2010fn} which would fit into a split supersymmetry scenario \cite{ArkaniHamed:2004fb, Giudice:2004tc}. Hence, hard wall models can actually cover all of GGM parameter space. In fact, it is not entirely satisfactory that a necessary ingredient for all this amounts to introduce two parameters, $H_{1/2}$ and $E_0$, which are directly proportional to gaugino and sfermions masses, respectively.  This is reminiscent of minimal benchmark points. It would thus be desirable to try and obtain both Majorana gaugino masses and positive squared sfermions masses by enriching the dynamics in the bulk instead of introducing inhomogeneous terms in the IR boundary conditions. In the next section we will achieve this goal by turning on a linear profile for an R-charged scalar, as it was done in \cite{Argurio:2012cd}. 

Let us finally mention that, as noticed in  \cite{Yanagida:2012ef}, a positive value for $C_1-C_0$ is a desirable feature, in that it helps raising 
the mass of the Higgs in gauge mediation scenarios. In our models,  this is
achieved by the same conditions which make the right hand side of
\eqref{erho} positive.

\section{Hard wall with R-symmetry breaking mode}
\label{Rsymmbreaking}

In this section we would like to construct a simple scenario in which
the R-symmetry is broken (and gaugino masses generated)
dynamically. We will follow the same logic as \cite{Argurio:2012cd},
where it was observed that  considering only  the minimal action
\eqref{minimalaction} for the vector supermultiplet it is impossible
to break the R-symmetry by bulk dynamics, and get non-zero gaugino
Majorana masses. As in the top-down model considered in
\cite{Argurio:2012cd}, we will see that the dynamical breaking of  R-symmetry implies automatically the absence of massless modes in $C_{1/2}$. Notice that this physical consistency condition had instead to be imposed by hand, in the previous section. 

We introduce a new dynamical scalar field $\eta$ in the bulk with $m^2=-3$, and treat it as a linear fluctuation around the hard wall metric.

The action for $\eta$ at the linearized level is completely determined by its mass while the precise values of its couplings with the vector multiplet can be guessed by analogy with the $N=2$ supergravity embedding considered in \cite{Argurio:2012cd},  based on the general results of \cite{Ceresole:2000jd,Ceresole:2001wi}\footnote{With respect to \cite{Argurio:2012cd} we have rotated the spinors by a phase in order to have a real $B_{1/2}$ at the end.}
\begin{align}
&S_{\text{kin}} = \frac{N^2}{4\pi^2}\int d^5x \sqrt{G}(G^{\mu\nu}\partial_{\mu}\eta\partial_{\nu}\eta-3\eta^2)\ ,\\
&S_{\text{int}}=\frac{N^2}{4\pi^2}\int d^5x\frac{\sqrt{G}}{2}\left[(\eta+z\partial_{z}\eta)(\chi\chi+\bar\chi\bar{\chi})+(\eta-z\partial_{z}\eta)(\xi\xi+\bar{\xi}\bar{\xi})\right]\ . \label{intEta}
\end{align}     
One might think that, in view of the possibility of constructing more general bottom-up models, it might be interesting to see what happens if we take arbitrary coefficients in the interactions term. On the other hand, asking for a gravity dual of a supersymmetric field theory (which then breaks supersymmetry spontaneously or by a soft deformation) puts severe constraints on the possible interactions. In fact,  precisely the constraints dictated by supergravity. One can check that choices other than the interactions above do not give the right supersymmetric result in the deep UV. 

We demand the R-symmetry breaking mode $\eta(z,x)$ to have a non-trivial profile in the vacuum which is independent on the boundary space-time directions in order to preserve Poincar\'e invariance of the boundary theory. The most general solution to the resulting equations of motion for $\eta$ without $k$ dependence is 
\begin{equation}
\eta(z)=z\eta_{0}+z^3\tilde{\eta}_{2}\ ,
\end{equation}
where $\eta_{0}$ and $\tilde{\eta}_{2}$ are two arbitrary constants. These two constants can be fixed imposing, as usual, boundary conditions at $z=0$ and at the IR cut-off $z=1/\mu$. This strictly amounts to considering them as free parameters, which we will do in the following.

The equations of motion for $\lambda$ are modified by the presence of the extra contribution (\ref{intEta}) and become 
\be
 \begin{cases}
(z\partial_{z}-\frac{5}{2})\chi+z\sigma^{i}k_{i}\bar{\xi}+(\eta-z\partial_{z}\eta)\xi=0\ ,\\
(-z\partial_{z}+\frac{3}{2})\bar{\xi}+z\bar{\sigma}^{i}k_{i}\chi+(\eta+z\partial_{z}\eta)\bar{\chi}=0\ .\label{fermionRbroken}
\end{cases}
\ee
Comparing with eqs.~(\ref{spinorminimalUV}), one can easily conclude that in the present case  the leading boundary behavior of the fermionic field is not modified with respect to the minimal case \eqref{boundary1/2}.  On the other hand, as discussed in \cite{Argurio:2012cd}, whenever $\eta_0 \not =0$, we have to modify the definition of the fermionic correlator defining $B_{1/2}$ according to
\be
\langle j_{\alpha}(k)j_{\beta}(-k)\rangle= \frac{\delta\tilde{\chi}_{1\alpha}}{\delta\xi_{0}^{\beta}}-\frac{\delta\tilde{\chi}_{1\beta}}{\delta\xi_{0}^{\alpha}}+2\eta_{0}\epsilon_{\alpha\beta} \ ,
\ee
while the expression for the non-chiral fermionic correlator \eqref{Ch1/2} remains unchanged. 

\begin{figure}
\begin{center}
\includegraphics[height=0.24\textheight]{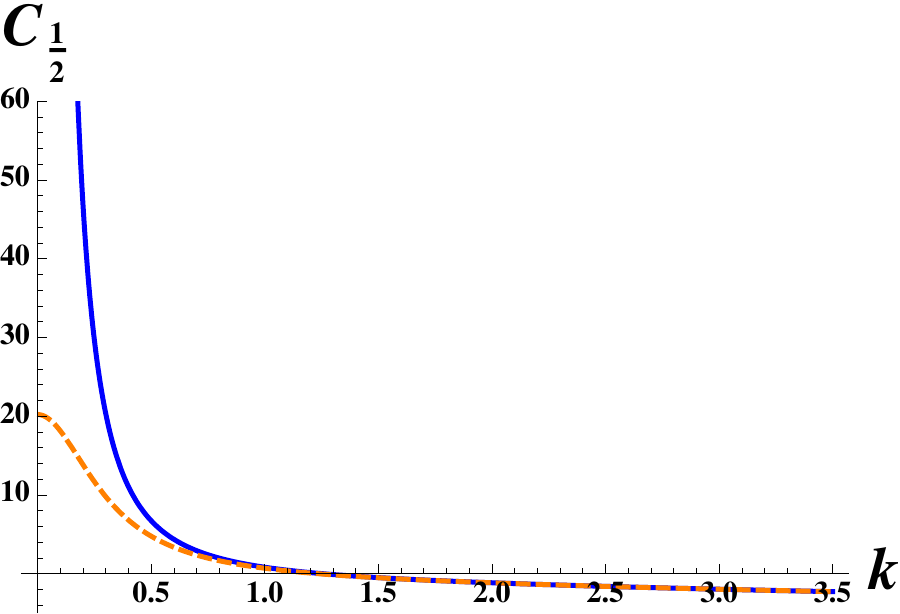}
\includegraphics[height=0.24\textheight]{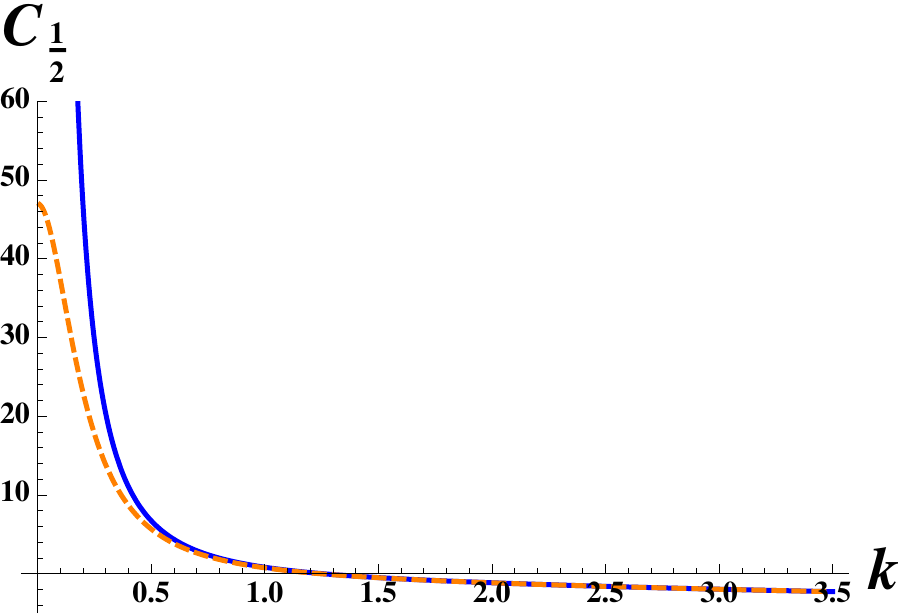}
\caption{\small The solid blue line is for $\eta = 0$, the dashed one on the left figure is for $\eta = 0.1 \ z^3$ and on the right figure for $\eta = 0.1 \ z$. In these plots $\rho_{1/2}=3$ and $\mu = 1$. Notice that turning on $\eta$ the massless pole disappears. \label{EtaC}}
\end{center}
\end{figure}

\begin{figure}
\begin{center}
\includegraphics[height=0.24\textheight]{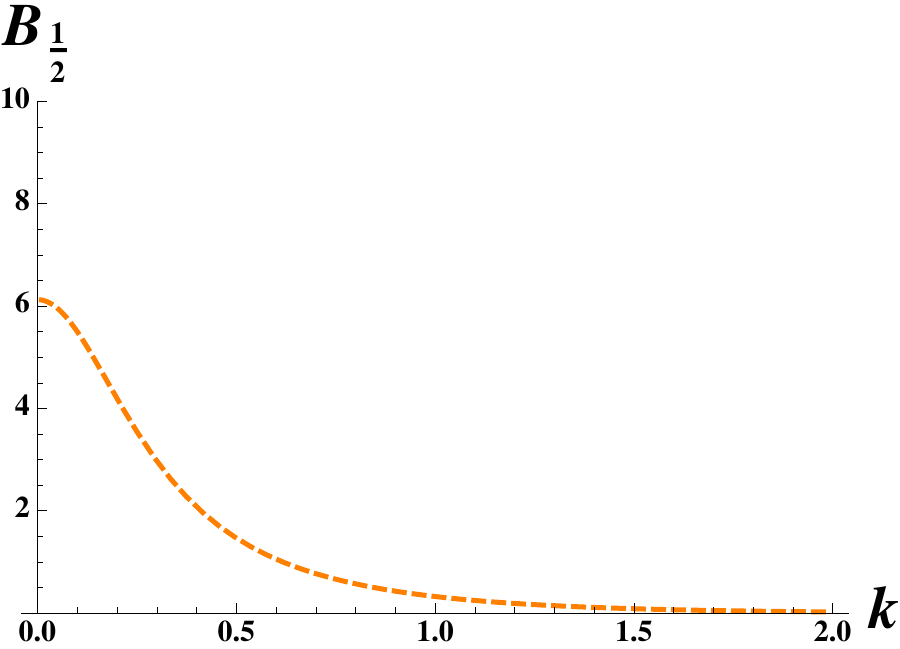}
\includegraphics[height=0.24\textheight]{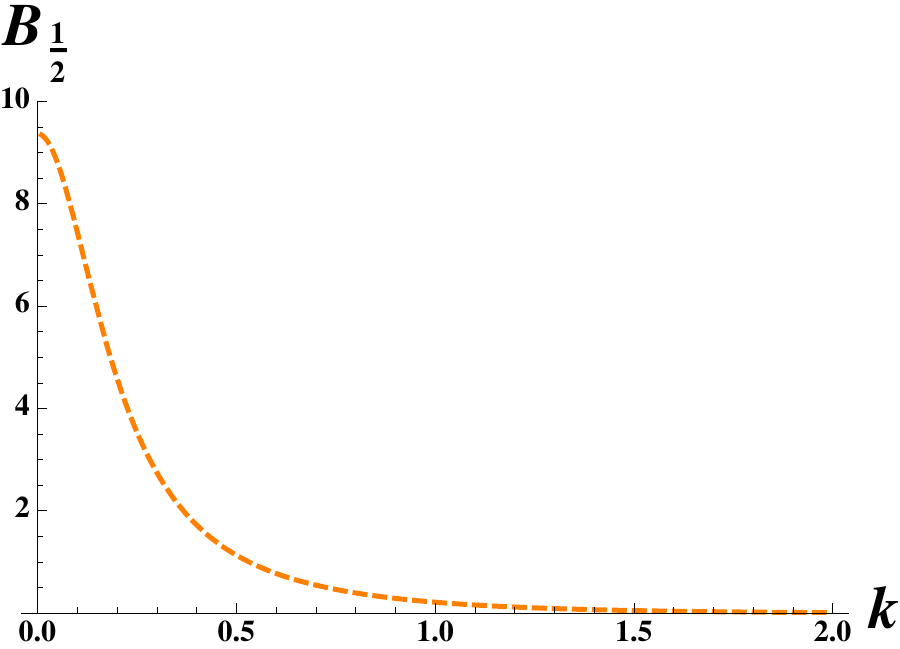}
\caption{\small On the left  $\eta = 0.1 \ z^3$, on the right  $\eta = 0.1 \ z$. In these plots $\rho_{1/2}=3$ and $\mu = 1$. When $\eta \neq 0$ a non-zero $B_{1/2}$ is generated. \label{EtaB}}
\end{center}
\end{figure}

We now need to solve eqs. \eqref{fermionRbroken} by imposing (homogeneous) boundary conditions in the IR (that for generic $\rho_{1/2}$ would give a massless pole when $\eta=0$). Unfortunately, this cannot be done analytically, and we have to resort to  numerics. Figures  \ref{EtaC} and  \ref{EtaB} contain our results. 

It is remarkable to see that when the R-symmetry is broken by a scalar
profile, the pole in $C_{1/2}$ disappears automatically.  
We note that the sfermion mass-squared is driven positive by
the fact that $C_{1/2}$ is still quite large near $k=0$, at least as
far as $\eta$ is a perturbation. If we stick to this model without
playing with inhomogeneous boundary conditions in the IR, it can be
seen that we are able to explore a smaller region of parameter space. (Possibly, a larger portion of parameter space can be reached by playing with $\rho_{1/2}$.)

While the above analysis is done numerically, it would be nicer to have some analytical control on (at least) the low momenta behavior of the correlators, to see, for instance,  how the pole in $C_{1/2}$ disappears when the R-charged scalar is turned on. This analysis turns out to be possible if we also take the parameters $\eta_0$ and $\tilde{\eta}_2$ parametrically small, and we obtain
\begin{align}
& C_{1/2} \simeq \frac{1+\frac 32 \rho_{1/2}}{1 + \frac 72 \rho_{1/2}} \ \dfrac{4\mu^2}{k^2 + 4 M_{\eta_0,\tilde{\eta}_2}^2} \label{EtaCIR} \ , \\
& B_{1/2} \simeq \frac{1+\frac 32 \rho_{1/2}}{1 + \frac 72 \rho_{1/2}}  \ \dfrac{8\mu^2 M_{\eta_0,\tilde{\eta}_2}}{k^2 + 4 M_{\eta_0,\tilde{\eta}_2}^2} \ ,\label{EtaBIR}
\end{align}
where
\be
M_{\eta_0,\tilde{\eta}_2}=\eta_0+\dfrac{1+\frac{11}{2}\rho_{1/2}}{1+\frac{7}{2}\rho_{1/2}}\dfrac{\tilde{\eta}_2}{\mu^2} \ .
\ee
Notice that these formulae agree with the numerical plots in figures \ref{EtaC} and \ref{EtaB}. The details of this computation are explained in appendix \ref{B}.

\section{Conclusions}

In this paper we have considered holographic models of gauge mediation, and used hard wall backgrounds as a prototype to see whether and how strongly coupled hidden sectors can actually cover the GGM parameter space.

For a generic choice of boundary conditions at the IR wall, the resulting low energy spectrum is that of mediation scenarios with extra non-SM gauge sectors, where $Z'$-like gauge bosons acquire a mass due to symmetry breaking in the hidden sector, and mediate supersymmetry breaking effects to the SM. Tuning some parameters one can eliminate the composite massless modes emerging in the hidden sector recovering more standard gauge mediation scenarios, and cover all of GGM parameter space. 

Hard wall models can be seen as an effective way to mimic non-trivial
bulk dynamics (for instance, several effects of a hard wall can be
obtained with a non-trivial dilaton profile). In this sense, it would
be interesting to see whether there exist more dynamical models where
part of all GGM parameter space can be reproduced. The analysis of
section 5 is an example of such a strategy. More ambitiously, it would
be interesting to see whether relaxing the AAdS-ness of the background (which, strictly speaking, makes the dual hidden sector dynamics somewhat unrealistic), changes any of the results.

On a more formal level, it might very well be that not all sets of physically distinct boundary conditions can find a counterpart in fully-fledged 10d string embeddings. This is a question which is very difficult to answer, but which certainly deserves further attention. 

We hope to return to these issues in the near future.

\section*{Acknowledgements}
We would like to thank Francesco Bigazzi, Aldo Cotrone, Stefano
Cremonesi, Alberto
Mariotti, Daniele Musso, Carlos Nunez,
Andrea Romanino, Marco Serone and Gary Shiu for useful discussions. 
The research of R.A. and D.R. is supported in part by IISN-Belgium (conventions
4.4511.06, 4.4505.86 and 4.4514.08), by the ``Communaut\'e
Fran\c{c}aise de Belgique" through the ARC program and by a ``Mandat d'Impulsion Scientifique" of the F.R.S.-FNRS. R.A. is a Research Associate of the Fonds de la Recherche Scientifique--F.N.R.S. (Belgium). M.B., L.D.P. and F.P. acknowledge partial financial support by the MIUR-PRIN contract 2009-KHZKRX.


\appendix

\section{Definition and properties of Bessel functions}
\label{A}

In this appendix we will briefly summarize the definitions and the basic properties of the Bessel functions following \cite{Bessel}. As we show in the paper, the equations for the fluctuations of the fields in the vector multiplet can be recast in 
the Bessel differential equation
\begin{equation}
(z^{2}\partial_{z}^{2}+z\partial_{z}-(z^2k^2+\nu^2))f(z,k)=0 \label{bessel}
\end{equation}
for $\nu=0,1$. Taking $k^2\geq0$, the two independent solutions of the above equation can be written as 
\begin{equation}
f(z,k)=c_{1}(k)I_{\nu}(zk)+c_{2}(k)K_{\nu}(zk)\ .
\end{equation}
The Bessel functions $I_{\nu}(x)$ and $K_{\nu}(x)$ can be defined as power series:
\begin{align}
&I_{\nu}(x)=\sum_{n=0}^{\infty}\frac{x^{2n+\nu}}{n!2^{\nu+2n}(\nu+n)!}\ , \label{defI}\\
&K_{\nu}(x)=\frac{1}{2}\sum_{n=0}^{\nu-1}\frac{(-)^{n}(\nu-n-1)!}{n!2^{2n-\nu}}x^{2n-\nu}+\notag\\
&\ \ \ \ \ \ \ \ \ \ \ (-)^{\nu+1}\sum_{n=0}^{\infty}\frac{x^{2n+\nu}}{n!2^{2n+\nu}(\nu+n)!}(\text{log}\left(\frac{x}{2}\right)-\frac{1}{2}\psi(n+1)-\frac{1}{2}\psi(\nu+n+1))\ ,\label{defK}
\end{align}
with 
\begin{equation}
\psi(n+1)=\sum_{m=1}^{n}\frac{1}{m}-\gamma\ ,
\end{equation}
where $\gamma\simeq 0.58$ is the Euler-Mascheroni constant.

Some recurrence formulas involving $I_{\nu}(x)$ and $K_{\nu}(x)$ are 
the following:
\begin{align}
&\partial_{x}I_{0}(x)=I_1(x), \nn\\
&\partial_{x}K_{0}(x)=-K_1(x), \nn\\
&x\partial_{x}I_{1}(x)+ I_{1}(x)=x I_0(x), \nn\\
&x\partial_{x}K_{1}(x)+ K_{1}(x)=-x K_0(x).
\label{recurrence}
\end{align}
We now list the asymptotic behavior of the Bessel functions of interest. 
In the limit $x\gg 1$ an analysis based on the integral representation of the Bessel functions shows that at leading order
\begin{align}
&I_{\nu}(x)\underset{\overset{{x\to\infty}}{}}{\simeq}\frac{e^{x}}{(2\pi x)^{1/2}}\ ,\label{largeI}\\
&K_{\nu}(x)\underset{\overset{{x\to\infty}}{}}{\simeq}\left(\frac{\pi}{2x}\right)^{1/2}e^{-x}\ ,\label{largeK}
\end{align}
independently on $\nu$.

In the limit $x\ll 1$ the Bessel functions are well approximated by the first terms of their series expansion (\ref{defI})--(\ref{defK}):
\begin{align}
&I_{0}(x)\underset{\overset{{x\to0}}{}}{\simeq}1+\frac{1}{4}x^2 +\dots, \nn\\
&I_{1}(x)\underset{\overset{{x\to0}}{}}{\simeq}\frac{1}{2}x+\frac{1}{16}x^3 +\dots, \label{Ismall}\\
&K_{0}(x)\underset{\overset{{x\to0}}{}}{\simeq}-\log x +\log 2 -\gamma +\dots,\nn\\
&K_{1}(x)\underset{\overset{{x\to0}}{}}{\simeq}\frac{1}{x}+\frac{1}{2}x\left(\log x -\log 2 -\frac{1}{2}+\gamma\right) +\dots.
\label{Ksmall}
\end{align}

\section{The IR limit of correlation functions}
\label{B}

In section 5 we have shown how the presence of a non-trivial profile for an R-charged scalar field, $\eta$, while providing a non-vanishing value for the R-breaking fermionic correlator $B_{1/2}$, consistently removes the pole from the non-chiral fermionic correlator $C_{1/2}$. The analysis was done by numerical methods. Here we show that one can actually study the IR behavior of holographic correlators analytically.

We are interested in analyzing the behavior of the correlation functions for small $k$. More precisely, the relevant quantity is $k/\mu \ll 1$, where $z=1/\mu$ is the position of the IR wall, so that the limit can also be seen as moving the wall closer to the boundary. This suggests that if we just need to evaluate the behavior of the $C_s$ functions at low momenta, i.e.~\eqref{hIR0}--\eqref{hIR1/2}, we can impose the IR boundary condition directly on the near-boundary expansion of the solutions, keeping only terms up to a mode high enough to match the order in $k^2$ at which we need the $C_s$. Indeed, in previous sections we have seen that this limit is very easy to obtain when one has exact solutions, since it involves expanding Bessel functions near the origin, i.e. keeping only the near-boundary expansion.

Let us illustrate this procedure with $C_0$ with homogenous IR boundary conditions.
We just need to substitute the expansion \eqref{boundary0} in the boundary conditions
\eqref{hwIR0}. We get
\be
\frac{1}{\mu^2}(d_0\log(\Lambda/\mu)+\tilde d_0 ) +\rho_0\frac{1}{\mu^2}(2d_0\log(\Lambda/\mu) +d_0 +2\tilde d_0)=0 \ ,
\ee
that is
\be
\tilde d_0 = -d_0 \left( \log(\Lambda/\mu) +\frac{\rho_0}{1+2\rho_0}\right) \ .
\ee
Applying
\be
C_0 = -  2\frac{\delta \tilde{d}_0}{\delta d_0} ,
\ee
we obtain \eqref{hIR0} right away and effortlessly. In order to reproduce eqs. \eqref{hIR1} and \eqref{hIR1/2}, the only added difficulty is that we have to go one order higher in the expansion,  if interested in both the $1/k^2$ pole and the finite term. 

Notice that this procedure works because the equations of motion themselves are not modified with respect to the AdS ones. If we had ${\cal O}(\mu)$ corrections to the metric (as in the example used in \cite{Argurio:2012cd}), it would be impossible to take $1/\mu$ small without introducing large corrections to the background metric and thus to the equations for the fluctuations. 

The case of an AdS hard wall with a scalar profile turned is a particular case. In order to prove that the pole in $C_{1/2}$ disappears when $\eta=\eta_0 z + \tilde{\eta}_2 z^3$ is turned on, we should take the limit $k \to 0$ in such a way to keep terms of the form $(k^2+\eta_0^2)^{-1}$ or $(k^2 + \mu^{-4}\tilde{\eta}_2^2)^{-1}$. Therefore, the correct scaling is
\be
 \eta_0/\mu \sim \tilde{\eta}_2/\mu^3 \sim k/\mu = \epsilon \to 0 \ ,
\ee
and we should focus on the order $\epsilon^{-2}$ in the small $\epsilon$ expansion of $C_{1/2}$. Keeping $\eta$ small we also ensure that we can still use the AdS near boundary expansion for the fluctuations. The same kind of expansion can be done for the $B_{1/2}$ correlator, with the difference that it starts from the $\epsilon^{-1}$ order. In both cases, the leading terms in the $\epsilon$ expansion receive a non-trivial contribution both from $\eta_0\neq 0$ and from $\tilde{\eta}_2\neq 0$ and they are determined by keeping the near-boundary expansion 
\be
\xi(z,x) = z^{3/2}\left[\xi_0 + \sum_{n=1}^{\infty}(\tilde{\xi}_{2n} + \xi_{2n} \log(z\Lambda))z^{2n}\right]
\ee
up to $n=1$ and $n=2$, respectively.
The results for the order $\epsilon^{-2}$ of $C_{1/2}$ and the order $\epsilon^{-1}$ in $B_{1/2}$ are reported in  section 5.
If one wants to go to the next order in $\epsilon$, which is order $\epsilon^0$ for $C_{1/2}$ and order $\epsilon$ for $B_{1/2}$, one should keep terms up to $n=3$ in the near-boundary expansion. Let us stress that this $\epsilon$ expansion is different from a simple expansion for small momenta. For instance, the finite $k = 0$ term will receive contribution from arbitrary high orders in $\epsilon$, which in turn would require to keep arbitrary high terms in the near boundary expansion. Nevertheless, as long as $\eta_0$ and $\tilde{\eta}_2$ are kept small, the approximations (\ref{EtaCIR})--(\ref{EtaBIR}) give a reliable information about the finite value at $k=0$, as can be checked with the numerical results plotted in figures \ref{EtaC} and \ref{EtaB}.


\bibliographystyle{plainnat}

\end{document}